\input harvmac 
\input pictex
\overfullrule=0pt
\def\half{{\textstyle{1\over 2}}}

\def\sech{\mathop{{\rm sech} }}
\def\arctanh{\mathop{{\rm arctanh} }}
\def\neqno#1{\eqnn{#1} \eqno #1}

\Title{hep-th/0009127, UTPT-00-11} {\vbox{
\centerline{The Cosmological Constant and Warped Extra Dimensions} }}
\centerline{\bf Hael Collins\footnote{$^\dagger$}{{\tt 
hael@physics.utoronto.ca}} and Bob 
Holdom\footnote{$^\ddagger$}{{\tt bob.holdom@utoronto.ca}} }
\medskip
\vbox{\it \centerline{Department of Physics} \vskip-2pt
\centerline{University of Toronto} \vskip-2pt
\centerline{Toronto, Ontario M5S 1A7, Canada}}

\bigskip
\bigskip
\centerline{Abstract}

\medskip 
{\baselineskip=10pt\centerline{\vbox{\hsize=.8\hsize\ninepoint\noindent  
We study the behavior of a general gravitational action, including quadratic
terms in the curvature, supplemented by a compact scalar field in $4+1$
dimensions.  The generalized Einstein equation for this system admits
solutions which are compact in one direction and Poincar\' e invariant in the
remaining directions.  These solutions do not require any fine-tuning of the
parameters in the action---including the cosmological constant---only that
they should satisfy some mild inequalities.  Some of these inequalities can be
expressed in a universal form that does not depend on the number of extra
compact dimensions when the scenario is generalized beyond $4+1$ dimensions.  
}}}

\Date{September, 2000}
\baselineskip=12pt

\newsec{Introduction.}

The old idea that the universe might contain more than the observed four
space-time dimensions has re-emerged recently in novel attempts to explain the
weakness of gravity compared to the other forces 
\ref\add{N.~Arkani-Hamed, S.~Dimopoulos and G.~Dvali, ``The hierarchy problem
and new dimensions at a millimeter,'' Phys.\ Lett.\  {\bf B429}, 263 (1998)
[hep-ph/9803315] and I.~Antoniadis, N.~Arkani-Hamed, S.~Dimopoulos and
G.~Dvali, ``New dimensions at a millimeter to a Fermi and superstrings at a
TeV,'' Phys.\ Lett.\  {\bf B436}, 257 (1998) [hep-ph/9804398].}  
and the hierarchy problem 
\ref\rsa{L.~Randall and R.~Sundrum, ``A large mass hierarchy from a small
extra dimension,'' Phys.\ Rev.\ Lett.\  {\bf 83}, 3370 (1999)
[hep-ph/9905221].},   
but it was realized earlier 
\ref\rubakov{V.~A.~Rubakov and M.~E.~Shaposhnikov, ``Extra Space-Time
Dimensions: Towards A Solution To The Cosmological Constant Problem,'' Phys.\
Lett.\  {\bf B125}, 139 (1983). }  
that such theories might be able to address the cosmological constant problem 
\ref\weinberg{S.~Weinberg, ``The Cosmological Constant Problem,'' Rev.\ Mod.\
Phys.\  {\bf 61}, 1 (1989).}.   
The hope is that with extra dimensions, the metric might be able, through a
non-trivial dependence on the extra coordinates, both to accommodate an
arbitrary value for the cosmological constant and to maintain Poincar\' e
invariance in $3+1$ of the directions.  In their original treatment, Rubakov
and Shaposhnikov $\rubakov$ found that this idea could be realized with two
extra dimensions but that the bulk metric contained a singularity.  Moreover,
their system was shown to be unstable by 
\ref\lt{G.~V.~Lavrelashvili and P.~G.~Tinyakov, ``On Possible Spontaneous
Compactification Leading To Zero Cosmological Constant,'' Sov.\ J.\ Nucl.\
Phys.\  {\bf 41}, 172 (1985).}.   
The more recent scenarios $\rsa$ involving $3$-branes reintroduce the
cosmological constant problem in the form of a fine-tuning between the bulk
cosmological constant and the brane tension.  The attempts by 
\ref\adks{N.~Arkani-Hamed, S.~Dimopoulos, N.~Kaloper and R.~Sundrum, ``A small
cosmological constant from a large extra dimension,'' Phys.\ Lett.\  {\bf
B480}, 193 (2000) [hep-th/0001197].}  
and
\ref\kss{S.~Kachru, M.~Schulz and E.~Silverstein, ``Bounds on curved domain
walls in 5d gravity,'' hep-th/0002121.}  
to avoid this fine-tuning again resulted in naked singularities in the bulk
\ref\uglyfeatures{See for example S.~S.~Gubser, ``Curvature singularities: The
good, the bad, and the naked,'' hep-th/0002160; 
C.~Csaki, J.~Erlich, C.~Grojean and T.~Hollowood, ``General properties of the
self-tuning domain wall approach to the cosmological constant problem,''
hep-th/0004133 or 
P.~Binetruy, J.~M.~Cline and C.~Grojean, ``Dynamical instability of brane
solutions with a self-tuning cosmological constant,'' hep-th/0007029.}.   

We pursue the original idea of Rubakov and Shaposhnikov---that a
non-factorizable metric that depends on an extra dimension could account for
the presence of a cosmological constant.  In particular, we seek metrics that
are periodic in this extra dimension so that it can be made naturally compact
when the period of the metric is identified with the compactification radius.

The standard Einstein equation for a theory without any $3$-branes but with a
cosmological constant does not permit metrics that simultaneously satisfy both
requirements---that the metric is periodic and free of singularities.  We
therefore consider the effect of including higher order terms in the
gravitational action which contain two powers of the curvature tensor 
\ref\wetterich{C.~Wetterich, ``Spontaneous Compactification In Higher
Dimensional Gravity,'' Phys.\ Lett.\  {\bf B113}, 377 (1982) and Q.~Shafi and
C.~Wetterich, ``Cosmology From Higher Dimensional Gravity,'' Phys.\ Lett.\ 
{\bf B129}, 387 (1983). },  
such as a Gauss-Bonnet term 
\ref\kkl{J.~E.~Kim, B.~Kyae and H.~M.~Lee, ``Effective Gauss-Bonnet
interaction in Randall-Sundrum compactification,'' Phys.\ Rev.\  {\bf D62},
045013 (2000) [hep-ph/9912344] and 
J.~E.~Kim, B.~Kyae and H.~M.~Lee, ``Various modified solutions of the
Randall-Sundrum model with the  Gauss-Bonnet interaction,'' Nucl.\ Phys.\ 
{\bf B582}, 296 (2000) [hep-th/0004005].}  
\ref\lowzee{I.~Low and A.~Zee, ``Naked singularity and Gauss-Bonnet term in
brane world scenarios,'' hep-th/0004124.}  
\ref\deruelle{N.~Deruelle and T.~Dolezel, ``Brane versus shell cosmologies in
Einstein and Einstein-Gauss-Bonnet theories,'' gr-qc/0004021.}.   
Such terms can be regarded as the next natural terms in an effective theory of
the gravitational action such as might arise in the low-energy expansion of
some quantum theory of gravity.  

We show that a general five-dimensional gravitational action including all
terms up to fourth order in derivatives when supplemented by a compact scalar
field permits metrics that are periodic in the fifth coordinate and preserve
Poincar\' e invariance in the other four dimensions.  This result does not
require any fine-tuning at the level of the parameters in the
action---including the cosmological constant---other than that they should
satisfy some mild bounds.  Moreover, since these solutions are smooth and
contain no singularities, either in the metric or in the scalar field, the
scenario does not require any $3$-branes with an accompanying requisite
fine-tuning between the brane tension and bulk cosmological constant.  The
only requirement is that the compactification radius should be sufficiently
small so as not to produce any discrepancies with current experiments.  

In the next section, we review the solution to Einstein's equations in a $5d$
theory with a cosmological constant and a free scalar field.  Section three
derives the field equations from a general action that includes quadratic
terms in the curvature.  In section four, we present several exact solutions
of these field equations and further show that with the Gauss-Bonnet term
alone the metric does not have periodic, smooth, non-singular solutions with
$4d$ Poincar\' e invariance.  The fifth section shows the requirements that a
general $R^2$ action should satisfy to admit periodic, smooth, non-singular
metrics and discusses a few representative cases found numerically.  Section
six examines a few general properties of including higher order terms in the
effective action and discusses theories with more than one extra dimension. 
Section seven concludes.

\newsec{Background.}

In order to show the importance of higher derivative terms in the action, we
first examine the solutions for a five-dimensional theory with only the
standard Einstein-Hilbert action and a free scalar field,\foot{Our convention
for the signature of the metric is $(-,+,+,+,+)$ while the Riemann curvature
tensor is defined by $-R^a_{\ bcd} \equiv \partial_d \Gamma^a_{bc} -
\partial_c \Gamma^a_{bd} + \Gamma^a_{ed}\Gamma^e_{bc} -
\Gamma^a_{ec}\Gamma^e_{bd}$.}
$$S = M_5^3 \int d^4xdy\, \sqrt{-g}\, \left( 2\Lambda  + R \right) 
- {1\over 2}\int d^4xdy\, \sqrt{-g}\, \left( \nabla_a\phi\nabla^a\phi\right) .
\neqno\simpleaction$$
Here $\Lambda$ and $M_5$ are respectively the cosmological constant and the
$5$ dimensional Planck constant.  $g_{ab}$ is the metric for the space-time. 
We denote the coordinates that correspond to the usual space-time dimensions
by $x^\mu$, where $\mu,\nu,\cdots = 0,1,2,3$, and the fifth coordinate by $y$,
with $a,b,c,\ldots=0,1,2,3,y$.  Ignoring the scalar field, when $\Lambda>0$
the universe is an anti-de Sitter (AdS) space-time while $\Lambda<0$
corresponds to a de Sitter (dS) space-time.  We shall often work in units in
which $M_5=1$.  

When the metric has the form
$$ds^2 = g_{ab}\, dx^a dx^b = e^{A(y)}\eta_{\mu\nu}\, dx^\mu dx^\nu + dy^2 ,
\neqno\rsmetric$$
the $\mu\mu$ and $yy$ components of the Einstein equation for $\simpleaction$
are
$$\eqalign{
- {\textstyle{3\over 2}} (A')^2 - {\textstyle{3\over 2}} A^{\prime\prime} &= -
\Lambda + {\textstyle{1\over 4}} (\phi')^2 \cr
- {\textstyle{3\over 2}} (A')^2 
&= - \Lambda - {\textstyle{1\over 4}} (\phi')^2  \cr}
\neqno\simpleeqns$$
while the equation for the scalar field is 
$$\phi^{\prime\prime} + 2 A' \phi' = 0 . \neqno\simplephieqn$$
A solution to equations $\simpleeqns$--$\simplephieqn$ is given by 
$$\eqalign{
e^{A(y)} &= e^{A_0} \left[ \cos\left( 2 \sqrt{-{\textstyle{2\over
3}}\Lambda}(y-y_0) \right) \right]^{1/2} \cr
\phi'(y) &= 2\sqrt{-\Lambda}\, \sec\left( 2 \sqrt{-{\textstyle{2\over
3}}\Lambda}(y-y_0) \right) \cr} , \neqno\withphi$$
where $A_0$ and $y_0$ are constants of integration, in addition to the trivial
$\phi'(y)=0$ solution.  When $\Lambda < 0$ (dS), we are able to obtain
periodic solutions.  However, the scalar field periodically becomes singular
and the metric becomes imaginary.  In order to remove this unacceptable
behavior from the space-time, we require $3$-branes to cut-off the manifold in
extra dimension before the ill-behaved region is encountered.  When
$\Lambda>0$ (AdS), only the $\phi'(y)=0$ solution is real and the theory can
only be compact if we return to the original Randall-Sundrum scenario $\rsa$. 
Thus for either sign of $\Lambda$, the scenario must contain some $3$-branes
which necessitates a fine-tuning of the brane tension with the value of the
cosmological constant.  

In the following, we shall see how the addition of a general $R^2$ action can
lead to an acceptable dependence of the warp function, $A(y)$, on an extra
compact dimension.  Note that such an action encounters a difficulty when the
extra dimension is not compact 
\ref\rsb{L.~Randall and R.~Sundrum, ``An alternative to compactification,''
Phys.\ Rev.\ Lett.\  {\bf 83}, 4690 (1999) [hep-th/9906064].},  
as noted in 
\ref\zk{Z.~Kakushadze, ``Localized (super)gravity and cosmological constant,''
hep-th/0005217 and ``Gravity in the Randall-Sundrum brane world revisited,''
hep-th/0008128.}.   
If we consider variations about a flat $4d$ metric, $\eta_{\mu\nu} \to
g^{(4)}_{\mu\nu}(x^\lambda)$, we can relate the five-dimensional curvature for
$\rsmetric$ to the four-dimensional curvature through
$$R(x^\mu,y) = e^{-A(y)}R^{(4)}(x^\mu) + \cdots ,\neqno\fivetofour$$
where $R^{(4)}$ is the curvature associated with $g_{\mu\nu}^{(4)}$.  By
integrating out the extra dimension, 
$$\int d^4xdy\, \sqrt{g}\, M_5^3 R(x^\mu,y) =
\int d^4x\sqrt{g^{(4)}} M_5^3 R^{(4)}(x^\mu) \int dy\, e^{A(y)} + \cdots ,
\neqno\integrateout$$
we determine the effective Planck's constant $M_4$ measured by a
four-dimensional observer in terms of the $5d$ Planck's constant $M_5$ through
$M_4^2 \equiv M_5^3 \int dy\, e^{A(y)}$.  As long as this integral is finite
we can define a four dimensional effective theory of gravity.  This argument
fails at the next order since 
$$\int d^4xdy\, \sqrt{g}\, R^2  =
\int d^4x\sqrt{g^{(4)}} \left( R^{(4)} \right)^2 \int dy\, 1 + \cdots ,
\neqno\integratertwo$$
so that in an effective theory, this term receives a correction proportional
to the volume of the extra dimension.  The problem only worsens at higher
orders where the $[R^{(4)}]^k$ terms receive an enhancement of $\int dy\, e^{
(2-k)A(y)}$.  One way $\zk$ to evade this difficulty occurs when the term
induced in the $4d$ effective theory is purely topological, as is the case for
the $4d$ Gauss-Bonnet term.  In this paper we instead consider a scenario with
compact extra dimensions along with a general $R^2$ action; a Gauss-Bonnet
term alone is insufficient for our picture, as we show below.

\newsec{Gravity from a Generalized Action.}

A generic action with up to four derivatives of metric can be written as  
$$S = S_\phi + M_5^3 \int d^4xdy\, \sqrt{-g}\, \left( 2\Lambda  + R + a\, R^2
+ b\, R_{ab} R^{ab} + c\, R_{abcd}R^{abcd} + \cdots \right) .
\neqno\action$$
The additional terms can be interpreted as a squared Weyl tensor, 
$$C_{abcd} C^{abcd} = {1\over 6} R^2 - {4\over 3} R_{ab}R^{ab} +
R_{abcd}R^{abcd} ,\neqno\weylsq$$
a Gauss-Bonnet term, 
$$E = R^2 - 4 R_{ab} R^{ab} + R_{abcd}R^{abcd}, \neqno\eulerterm$$
and a third possible independent term,
$${\cal T} = {1\over 2} R^2 - 3 R_{ab}R^{ab} + R_{abcd}R^{abcd} .
\neqno\thirdterm$$
The Weyl term vanishes when the metric is conformally flat, as is the case for
$\rsmetric$, so we are free to add some multiple of the Weyl term to
$\thirdterm$ without affecting the field equations.  Later we shall use $\mu$
and $\lambda$ to denote the coefficients of the third term $\thirdterm$ and
Gauss-Bonnet term respectively:
$$\mu \equiv 16a + 5b + 4c \qquad\qquad 3\lambda \equiv 10a + 2b + c .
\neqno\mldefs$$

We include the effect of a scalar field, initially through the action for a
free field,
$$S_\phi = \int d^4xdy\, \sqrt{-g}\, \left( - \half k_0 \nabla_a\phi
\nabla^a\phi \right)  , \neqno\phiaction$$
although later we shall add a term $(\nabla_a\phi\nabla^a\phi)^2$ which is of
the same order as the $R^2$ terms.  Note that we have included a coefficient
$k_0$ in equation $\phiaction$.  

Upon varying the action with respect to the metric we obtain 
\ref\deWitt{B.~S.~DeWitt, ``Dynamical Theory of Groups and Fields,''
{\it  New York:  Gordon and Breach\/} (1965).}
$$\eqalign{
-g_{ab}\Lambda + R_{ab} - \half g_{ab}R - \half g_{ab}\, \left[ a\, R^2 + b\,
R_{cd}R^{cd} + R_{cdef}R^{cdef} \right] & \cr
+2a\, RR_{ab} - 4c\, R_{ac}R^c_{\ b} + 2c\, R_{acdf}R_b^{\ cdf} - 2(b+2c)\,
R^{cd}R_{acdb} & \cr
+\half(4a+b)\, g_{ab} \nabla^2 R - (2a+b+2c)\, \nabla_a\nabla_b R + (b+4c)\,
\nabla^2 R_{ab} &= T_{ab} \cr}
\neqno\eom$$
where $T_{ab}$ represents the energy-momentum tensor for the scalar field,
$$T_{ab} \equiv \half k_0 \left[ \nabla_a\phi\nabla_b\phi 
- \half g_{ab}\nabla_c\phi\nabla^c\phi \right] .
\neqno\tdef$$
For a non-factorizable metric of the form of equation $\rsmetric$, the sum of
the $\mu\mu$ and the $yy$ components of $\eom$ yields an equation in which the
free scalar field does not appear,
$$\eqalign{
\mu\left[ \half A^{\prime\prime\prime\prime} + 3 A' A^{\prime\prime\prime} +
(A^{\prime\prime})^2 + 4 A^{\prime\prime} (A')^2 \right] \qquad & \cr
+ {\textstyle{3\over 2}} \lambda \left[ A^{\prime\prime} (A')^2 + (A')^4
\right] - 3 (A')^2 - {\textstyle{3\over 2}} A^{\prime\prime} &= - 2 \Lambda ,
\cr}
\neqno\aeqn$$
so that the difference of these components can be used to determine its
behavior,
$$k_0(\phi')^2 = \mu\left[ A^{\prime\prime\prime\prime} + 2 A'
A^{\prime\prime\prime} + 4 (A^{\prime\prime})^2 \right] + 3 \lambda
A^{\prime\prime} (A')^2 - 3 A^{\prime\prime} .
\neqno\phiequal$$

The scalar field equation appears in $\simplephieqn$, but it is not
independent being a consequence of $\aeqn$ and $\phiequal$.  When $A(y)$ is a
periodic function of $y$, then $\phiequal$ implies that $\phi'(y)$ should also
be periodic.  Since this equation is non-linear in $A(y)$, the integral of
$\phi'(y)$ over one period is in general finite and non-zero so $\phi$ must
itself be compact.  Any additional dependence of the action on $\phi$, rather
than on its derivatives, must be through periodic functions.

\newsec{Analytical Solutions.}

The full fourth-order set of differential equations for a theory with a free
scalar field and a set of $R^2$ terms with arbitrary coefficients does not
admit a simple analytic solution except for special cases.  In this section,
we study two such examples of exact solutions.  Although these solutions do
not produce warp functions $A(y)$ that are smooth, non-singular and periodic,
they provide partial boundaries for the region of the $\{\Lambda, \lambda,
\mu\}$ parameter space in which we have found such solutions numerically.  One
of these surfaces requires both the $\mu$ and $\lambda$ terms in the field
equations.  Among this class of solutions is a metric which falls off
exponentially as $y\to\pm\infty$, as in $\rsb$ but without the need of a
$3$-brane.  A second of the surfaces, $\mu=0$, represents a theory with only a
Gauss-Bonnet term at the $R^2$ order.  Note that a third boundary lies along
the surface $\Lambda=0$ where we can trivially satisfy the field equations
$\aeqn$, $\phiequal$ and $\simplephieqn$ with constant solutions, $A(y)=A_0$
and $\phi(y)=\phi_0$.  We have found non-trivial periodic solutions only when
$\Lambda<0$ and $\mu<0$.

\subsec{An Exact Solution.}

We can discover an interesting set of exact solutions by noting that the
linear combination of field equations that eliminates the scalar field,
$\aeqn$, does not depend on $A(y)$ except through its derivatives and does not
contain the fifth coordinate explicitly.  Together these properties allow
$\aeqn$ to be recast as a second order differential equation,
$$\eqalign{
\mu \left[ {1\over 2} P^2 {d^2 P\over dz^2} + {1\over 2} P \left( {dP\over dz}
\right)^2 + 3 z P {dP\over dz} + P^2 + 4P z^2 \right] \quad & \cr
+ {3\over 2} \lambda \left[ Pz^2 + z^4 \right] - 3z^2 - {3\over 2} P &= -
2\Lambda , \cr} \neqno\zeqn$$
through the introduction of 
$$z \equiv {dA\over dy} \qquad\hbox{and}\qquad P(z) \equiv {dz\over dy} .
\neqno\zdefs$$
A simple set of solutions in the full $\{\Lambda,\lambda,\mu\}$ parameter
space is found by substituting 
$$P(z) = a z^2 + b \neqno\parabolicsoln$$
into $\zeqn$ which imposes the constraints
$$\Lambda = {3\over 4} {b\over a+2} \qquad 
\lambda = - {a\over b} {3a+4\over a+2} \qquad
\mu = {3\over 2} {1\over b} {1\over a+2} . \neqno\spsolnc$$
The resulting warp function $A(y)$ is periodic provided that $ab>0$:
$$e^{A(y)} = e^{A_0} \left[ \cos\left( \sqrt{ab}(y-y_0) \right) \right]^{-1/a}
. \neqno\expspsoln$$
Note that this solution for $A(y)$, when substituted into $\phiequal$, implies
that the scalar field is constant, $\phi'(y)=0$.

In general, this solution is not satisfactory since it contains singularities
when $a>0$ and becomes complex for generic values of $a<0$.  When $a=-1/2n$,
where $n$ is a positive integer, the behavior improves so that $e^{A(y)}$ is
everywhere real and non-singular when $\Lambda<0$, although for this case
$e^{A(y)}$ vanishes at regular intervals.  We shall find in the next section
that the set of solutions in $\expspsoln$ forms a boundary in the 
$\{\Lambda,\lambda,\mu\}$ parameter space beyond which periodic, non-vanishing
solutions exist when $\Lambda<0$ and $\mu<0$. 

We can also use $\expspsoln$ to generate a variant of the usual
Randall-Sundrum scenario $\rsb$, but without including a $3$-brane.  For
example, when $a>0$ and $\Lambda<0$, we find that 
$$e^{A(y)} = e^{A_0} \sech\left( \sqrt{-{\textstyle{4\over 3}} a(a+2)
\Lambda}\, (y-y_0) \right)^{1/a} . \neqno\exprstype$$
Although this solution requires fine-tuning the parameters in the action, as
given in $\spsolnc$, it achieves a metric whose zero mode is centered about
$y=y_0$ and which falls off exponentially as $y\to\pm\infty$.  Unlike previous
examples 
\ref\gremm{M.~Gremm, ``Four-dimensional gravity on a thick domain wall,''
Phys.\ Lett.\  {\bf B478}, 434 (2000) [hep-th/9912060].}  
which used a thick brane to obtain this behavior, $\exprstype$ does not
require any scalar field.  Since this theory has an infinite extra dimension,
some method of trapping the Standard Model fields near $y=y_0$ is further
required.

The endpoints of the set of solutions in $\exprstype$, shown in figure 1, have
interesting properties.  Taking $a\to\infty$ ($\mu=0$,
$\lambda\Lambda=-{9\over 4}$), so that the region in which the first
derivative of the warp function smoothly changes sign has a vanishing
thickness, the solution $\exprstype$ approaches the warp function studied by
Randall and Sundrum $\rsb$,
$$e^{A(y)} = e^{A_0} e^{- 2\sqrt{ {-\Lambda\over 3} } | y-y_0 | }.
\neqno\pseudors$$
At the other endpoint, $a\to 0$ ($\lambda=0$, $\mu\Lambda={9\over 32}$), the
warp function produces a Gaussian metric
$$e^{A(y)} = e^{A_0} e^{{4\Lambda\over 3}\, (y-y_0)^2} , \neqno\gaussian$$
which for $\Lambda<0$ decreases more rapidly away from $y=y_0$ than the
standard Randall-Sundrum metric.

\subsec{Insufficiency of a Gauss-Bonnet Term Alone.}

A theory with only a Gauss-Bonnet contribution to the $R^2$ action has the
advantage of producing a set of differential equations with no more than
second derivatives of the warp function $A(y)$.  Unfortunately, for our
purpose restricting to such a theory too greatly constrains the form of the
solutions.  Without the $\mu$-terms, equation $\aeqn$ implies that all the
extrema of $A(y)$, if they exist, are all local maxima or all local minima
depending upon the sign of $\Lambda$ since at the points where $A'(y)=0$,
$A^{\prime\prime}(y) = {4\over 3} \Lambda$.  However, the surface $\mu=0$ is
still important as it provides a partial boundary to the region in the full
$\{ \Lambda,\lambda,\mu \}$ parameter space in which satisfactory, periodic
solutions do exist.

When the metric has the form $\rsmetric$, any warp function $A(y)$ that has an
extremum will encounter a singularity only a finite distance away from it. 
Integrating the scalar field equation $\simplephieqn$, $\phi'(y) =
\sqrt{-4\Lambda}\, e^{-2(A(y)-A_0)}$,\foot{With $\phi'(y)=0$, the only
possible solutions are those of the form,
$$A(y) = \pm \left[ 1\pm \sqrt{1-{\textstyle{4\over 3}}\lambda \Lambda}
\right]^{1/2}\, {y-y_0\over\sqrt{\lambda}} + A_0 \neqno\nophinomu$$ 
where any of the four possible sign choices is allowed.  The case of a scalar
field in a $\Lambda=0$ theory was solved in $\lowzee$. } 
 and substituting this result into the linear combination of $\aeqn$ and
$\phiequal$ that eliminates $A^{\prime\prime}(y)$ when $\mu=0$ yields a first
order differential equation.  Its solution determines $A(y)$ implicitly:  
$$\eqalign{ \pm (y-y_0)
&= {\textstyle{ {\sqrt{\lambda}\over 2\sqrt{x_0+1}} }}
          \arctanh \left( {\textstyle{ \sqrt{ {1\pm x\over x_0+1}} }}\right) 
 - {\textstyle{ {\sqrt{\lambda}\over 2\sqrt{x_0-1}} }}
          \arctan \left( {\textstyle{ \sqrt{ {1\pm x\over x_0-1}} }}\right) 
 \cr} \neqno\nomusoln$$
where  
$$x\equiv\sqrt{x_0^2 + {\textstyle{4\over 3}}\lambda\Lambda e^{-4(A(y)-A_0)} }
\qquad\hbox{and}\qquad 
x_0 \equiv \sqrt{ 1-{\textstyle{4\over 3}}\lambda\Lambda } . \neqno\adefined$$
The signs on the right side of $\nomusoln$ cannot be chosen separately;
however, the minus sign should be used to obtain $A'(y)=0$ at some point.  In
the limit $\lambda\to 0$, $\nomusoln$ reproduces $\withphi$.  

In the interesting $\Lambda<0$ region, from the expression $\nomusoln$ we
observe that no acceptable solutions exist for any value of $\lambda$.  For
example when $\lambda<0$, $A(y)$ becomes singular in $\nomusoln$ at a finite
distance from $y_0$.  For $\lambda>0$, the warp function also contains a
singularity, but this time it appears in its second derivative.  Solving
equation $\aeqn$ for $A^{\prime\prime}(y)$, we note that it becomes singular
whenever $A'(y)=\pm\lambda^{-1/2}$, which again occurs at a finite value of
$y$.  Appendix A contains a fuller discussion of the properties of the
solution $\nomusoln$ and the location of its singularities.

\newsec{An Analysis of the Full $R^2$ Action.}

Before describing our approach for finding numerical solutions and presenting
several representative examples, we summarize with a sketch of the parameter
space in figure 1.  The unshaded area of the figure, which lies in the
$\Lambda<0$ and $\mu <0$ region of the $\{ \Lambda, \lambda, \mu \}$ parameter
space, shows where smooth, non-singular, periodic warp functions exist.  We
can describe another boundary by solving for $\lambda$ as a function of $\mu$
in $\spsolnc$, 
$$\lambda\Lambda = - {8\over 3}\, \mu\Lambda \pm 4 \sqrt{2\mu\Lambda} -
{9\over 4}; \neqno\spsolncmore$$
we immediately observe that the shape of the curve of solutions $\expspsoln$
is the same for all values of the cosmological constant, $\Lambda<0$, if the
$\lambda$ and $\mu$ axes are appropriately rescaled.  This feature corresponds
to a rescaling invariance of the field equation $\aeqn$ under $y\to\sigma y$,
$\mu\to\sigma^2\mu$, $\lambda\to\sigma^2\lambda$ and
$\Lambda\to\sigma^{-2}\Lambda$, where $\sigma$ is a real constant.  Therefore,
we can express our results in terms of two dimensionless parameters, $\{
\lambda\Lambda, \mu\Lambda \}$.
$$\beginpicture
\setcoordinatesystem units <0.5truein,0.5truein>
\setplotarea x from -3.0 to 5.0, y from -4.0 to 1.0
\putrule from -3.0 0.0 to 5.0 0.0
\putrule from 0.0 -4.0 to 0.0 1.0
\putrule from -2.0 -0.07 to -2.0 0.07
\putrule from -1.0 -0.07 to -1.0 0.07
\putrule from 1.0 -0.07 to 1.0 0.07
\putrule from 2.0 -0.07 to 2.0 0.07
\putrule from 3.0 -0.07 to 3.0 0.07
\putrule from 4.0 -0.07 to 4.0 0.07
\putrule from -0.07 -3.0 to 0.07 -3.0
\putrule from -0.07 -2.0 to 0.07 -2.0
\putrule from -0.07 -1.0 to 0.07 -1.0
\putrule from -0.07 1.0 to 0.07 1.0
\put {$\lambda$} [c] at 5.15 0.30 
\put {$\mu$} [c] at 0.30 1.15 
\put {$0$} [c] at -0.30 -0.20 
\put {$1$} [c] at -0.30 1.15 
\put {$1$} [c] at 1.00 -0.30 
\setshadegrid span <2.0pt>
\vshade -3 0 1  5 0 1 /
\hshade -4 -3 5  <,,z,z>  -3.375 -3 5 <,,z,z> 
-.6328 -.5625 5 <,,z,z>  -.60 -.5316 5 <,,z,z> 
-.40 -.2610 5 <,,z,z>  -.30 -.0484 5 <,,z,z> 
-.25 .0883 5 <,,z,z>  -.225 .1667 5 <,,z,z>  
-.167  .3842 5  <,,z,z>  -.150 .4591 5 <,,z,z>  
-.100 .7279 5 <,,z,z>  -.050 1.1184 5 <,,z,z>  
0.000 2.2500 5  /
\setlinear
\plot 1.603 -4.00  1.479 -3.90   1.356 -3.80  1.236 -3.70
1.117 -3.60  1.000 -3.50  .886 -3.40  .774 -3.30
.664 -3.20  .5570 -3.10  .4520 -3.00  .3501 -2.90
.2511 -2.80  .1548 -2.70  .0617 -2.60  -.0277 -2.50
-.1136 -2.40  -.1959 -2.30  -.2737 -2.20  -.3476 -2.10
-.4167 -2.00  -.4809 -1.90  -.5396 -1.80  -.5923 -1.70
-.6385 -1.60  -.6780 -1.50  -.7099 -1.40  -.7329 -1.30
-.7468 -1.20  -.7495 -1.10  -.7401 -1.00  -.7164 -.90
-.6763 -.80  -.6161 -.70  -.5316 -.60  -.4167 -.50
-.2610 -.40  -.0484 -.30  0 -.28125  .0883 -.25  .1667 -.225 
.2535 -.200  .3503 -.175  .4591 -.150  .5833 -.125 
.7279 -.100  .9008 -.075  1.1184 -.050  1.4223 -.025 
2.2500 0.00  3.2111 -.025  3.6482 -.050  3.9992 -.075 
4.3055 -.100  4.5833 -.125  4.8409 -.150  5.0000 -.167  /
\plot  0 -.26125  .0883 -.230  .1667 -.205 
.2535 -.180  .3503 -.155  .4591 -.130  .5833 -.105 
.7279 -.080  .9008 -.055  1.1184 -.030  1.4223 -.005 
2.2500 0.02  /
\plot  0 -.27125  .0883 -.240  .1667 -.215 
.2535 -.190  .3503 -.165  .4591 -.140  .5833 -.115 
.7279 -.090  .9008 -.065  1.1184 -.040  1.4223 -.015 
2.2500 0.01  /
\plot  0 -.29125  .0883 -.260  .1667 -.235 
.2535 -.210  .3503 -.185  .4591 -.160  .5833 -.135 
.7279 -.110  .9008 -.085  1.1184 -.060  1.4223 -.035 
2.2500 -0.01  /
\plot  0 -.30125  .0883 -.270  .1667 -.245 
.2535 -.220  .3503 -.195  .4591 -.170  .5833 -.145 
.7279 -.120  .9008 -.095  1.1184 -.070  1.4223 -.045 
2.2500 -0.02  /
\endpicture$$
{\ninepoint  \baselineskip=10pt
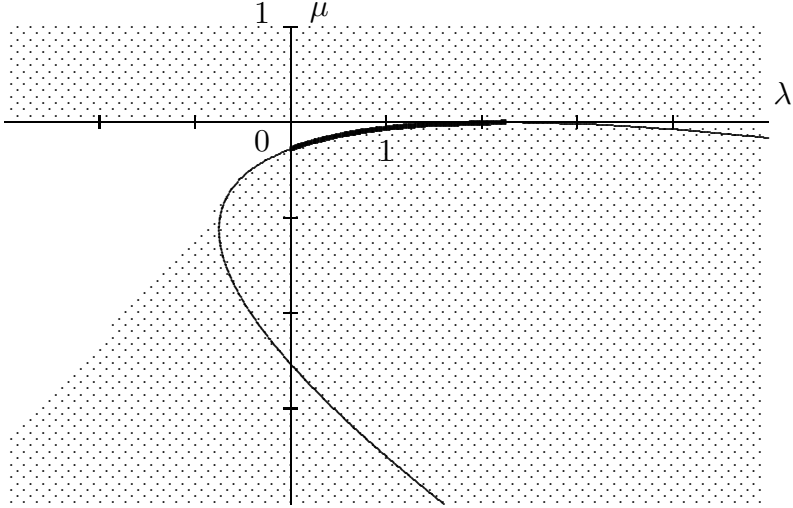
{\bf Figure 1.\/}  A plot of the parameter space $\{ \lambda, \mu \}$.  For
convenience, we have chosen $\Lambda=-1$ in generating this figure.  The
shaded region does not appear to contain periodic solutions for $A(y)$ while
in the unshaded region, we have found periodic solutions numerically for
arbitrarily chosen points.  The curve depicts the surface of solutions of
equation $\expspsoln$; the darker line shows the location of the solutions in
$\exprstype$.  }
\medskip

A periodic solution for a generic set of values of $\Lambda$, $\lambda$ and
$\mu$ is found by numerically integrating the differential equation $\aeqn$. 
The coordinate $y$ does not explicitly appear in the equation $\aeqn$ which
moreover only depends on the warp function through its derivatives.  Thus, we
can always translate by $y\to y-y_0$ and $A(y)\to A(y)+A_0$ to obtain another
solution.  Therefore we can choose $A(0) = A'(0) =0$ without any loss of
generality.  We also chose $A^{\prime\prime\prime}(0) = 0$ which limits our
solutions to those that are even about the origin, although we did not find
any periodic solutions that are odd when we relaxed this constraint.  The
subsequent evolution of the warp function away from $y=0$ then depended solely
upon the initial value of the second derivative, $A^{\prime\prime}(0)$.  

An arbitrary value for $A^{\prime\prime}(0)$, given some set of values for
$\{\lambda\Lambda, \mu\Lambda\}$, does not lead to a periodic solution. 
Generically, the warp function tended to reach a singularity at a finite
distance or to approach asymptotically a solution of the form
$A'(y)=\hbox{constant}$ as $y\to\infty$.  Between these two extremes,
precisely chosen values of $A^{\prime\prime}(0)$ produce periodic solutions
for arbitrarily chosen values of $\lambda\Lambda$ and $\mu\Lambda$ within the
unshaded region of figure 1.  This result demonstrates the existence of
periodic solutions which naturally select a compactification radius without
the need to add a $3$-brane to the theory or to fine-tune the parameters in
the action $\action$, as long as they lie within the allowed region of
parameter space.  

When the scalar field enters the action only through a free kinetic term, the
value of $k_0(\phi')^2$ is negative for all the periodic solutions that we
found numerically. Although this result implies that the kinetic energy term
has the wrong sign, it appears to be an artifact of truncating the scalar
action; this unphysical feature disappears when higher order terms appear in
the scalar action.  An example of such a solution is placed at the end of this
section.  

One region of parameter space that admits a semi-analytic approximation of the
solution is where $\Lambda\mu\sim\epsilon^2\ll 1$ with $\Lambda\lambda\not\gg
1$.  In this region we can try an oscillating solution for the warp function
of the form
$$A(y) = A_0 + \epsilon\cos(\omega (y-y_0)) + A_2 \epsilon^2\cos(2\omega
(y-y_0)) + {\cal O}(\epsilon^3) \neqno\cosapprox$$
for a small amplitude, $\epsilon$.  When substituted into $\aeqn$, the only
terms that are ${\cal O}(\epsilon)$ are those linear in $A(y)$:
$$\half \mu A^{\prime\prime\prime\prime} - {\textstyle{3\over 2}}
A^{\prime\prime} = {\cal O}(\epsilon^2) \neqno\aeqnapprox$$
which, to leading order, requires $\omega^2 \approx -3/\mu$ so that the period
of the compact dimension is
$$y_c \approx 2\pi\sqrt{-\mu/3} .\neqno\critradest$$
The ${\cal O}(\epsilon^2)$ terms, which arise from the following terms in
$\aeqn$,
$$\half\mu A^{\prime\prime\prime\prime} + 3\mu A' A^{\prime\prime\prime} + \mu
(A^{\prime\prime})^2 - 3 (A')^2 - {\textstyle{3\over 2}} A^{\prime\prime} = -
2 \Lambda + {\cal O}(\epsilon^3) , \neqno\aeqnsecond$$
determine the value of $\epsilon$ in terms of the cosmological constant, 
$$\epsilon^2 \approx - {4\over 3}{\Lambda\over\omega^2} ,
\neqno\leadingresultb$$
and the coefficient of the ${\cal O}(\epsilon^2)$ term in the warp function: 
$A_2 = - {1\over 4}$.  Thus, for the linear terms to dominate requires that
$\Lambda\mu \approx {9\over 4}\epsilon^2 \ll 1$.  Note that since the
$\lambda$ terms in the field equations contain at least three powers of the
warp function, to this order $\lambda$ can be arbitrary provided it does not
alter the $\epsilon$-expansion:  $\Lambda\lambda\not\gg 1$.

This semi-analytic estimate is confirmed when we plot the numerical solution
to $\aeqn$ for $\Lambda=-1$, $\lambda=0$ and $\mu=-0.01$ shown in figure 2. 
Holding $\Lambda=-1$ and $\lambda=0$ fixed and varying $\mu$, we have checked
that the period and amplitude of the numerical solutions is better and better
approximated by $\critradest$ and $\leadingresultb$ as we let $\mu\to 0$.  
$$\beginpicture
\setcoordinatesystem units <1.0truein,1.0truein>
\setplotarea x from 0.0 to 3.3, y from -0.1 to 1.4
\setlinear
\plot  0.000 0.000  0.015 0.003  0.030 0.011  0.045 0.024
0.060 0.042  0.075 0.066  0.090 0.094  0.105 0.127
0.120 0.164  0.135 0.205  0.150 0.249  0.165 0.297
0.180 0.347  0.195 0.400  0.210 0.455  0.225 0.511
0.240 0.568  0.255 0.626  0.270 0.684  0.285 0.742
0.300 0.800  0.315 0.856  0.330 0.911  0.345 0.964
0.360 1.014  0.375 1.063  0.390 1.108  0.405 1.150
0.420 1.189  0.435 1.224  0.450 1.255  0.465 1.282
0.480 1.305  0.495 1.323  0.510 1.337  0.525 1.346
0.540 1.351  0.555 1.351  0.570 1.346  0.585 1.336
0.600 1.322  0.615 1.304  0.630 1.281  0.645 1.254
0.660 1.223  0.675 1.187  0.690 1.149  0.705 1.106
0.720 1.061  0.735 1.013  0.750 0.962  0.765 0.909
0.780 0.854  0.795 0.798  0.810 0.740  0.825 0.682
0.840 0.624  0.855 0.566  0.870 0.509  0.885 0.452
0.900 0.398  0.915 0.345  0.930 0.295  0.945 0.247
0.960 0.203  0.975 0.162  0.990 0.125  1.005 0.093
1.020 0.065  1.035 0.042  1.050 0.023  1.065 0.010
1.080 0.002  1.095 0.000  1.110 0.003  1.125 0.011
1.140 0.025  1.155 0.043  1.170 0.067  1.185 0.095
1.200 0.128  1.215 0.165  1.230 0.206  1.245 0.251
1.260 0.298  1.275 0.349  1.290 0.402  1.305 0.457
1.320 0.513  1.335 0.570  1.350 0.628  1.365 0.687
1.380 0.745  1.395 0.802  1.410 0.858  1.425 0.913
1.440 0.966  1.455 1.016  1.470 1.064  1.485 1.110
1.500 1.152  1.515 1.190  1.530 1.225  1.545 1.256
1.560 1.283  1.575 1.305  1.590 1.324  1.605 1.337
1.620 1.346  1.635 1.351  1.650 1.351  1.665 1.346
1.680 1.336  1.695 1.322  1.710 1.303  1.725 1.280
1.740 1.253  1.755 1.221  1.770 1.186  1.785 1.147
1.800 1.105  1.815 1.059  1.830 1.011  1.845 0.960
1.860 0.907  1.875 0.852  1.890 0.795  1.905 0.738
1.920 0.680  1.935 0.622  1.950 0.564  1.965 0.507
1.980 0.450  1.995 0.396  2.010 0.343  2.025 0.293
2.040 0.246  2.055 0.201  2.070 0.161  2.085 0.124
2.100 0.092  2.115 0.064  2.130 0.041  2.145 0.023
2.160 0.010  2.175 0.002  2.190 0.000  2.205 0.003
2.220 0.012  2.235 0.025  2.250 0.044  2.265 0.068
2.280 0.096  2.295 0.129  2.310 0.167  2.325 0.208
2.340 0.252  2.355 0.300  2.370 0.351  2.385 0.404
2.400 0.459  2.415 0.515  2.430 0.573  2.445 0.631
2.460 0.689  2.475 0.747  2.490 0.804  2.505 0.860
2.520 0.915  2.535 0.968  2.550 1.018  2.565 1.066
2.580 1.111  2.595 1.153  2.610 1.192  2.625 1.226
2.640 1.257  2.655 1.284  2.670 1.306  2.685 1.324
2.700 1.338  2.715 1.347  2.730 1.351  2.745 1.350
2.760 1.345  2.775 1.336  2.790 1.321  2.805 1.302
2.820 1.279  2.835 1.252  2.850 1.220  2.865 1.185
2.880 1.145  2.895 1.103  2.910 1.057  2.925 1.009
2.940 0.958  2.955 0.905  2.970 0.850  2.985 0.793
3.000 0.736  3.015 0.678  3.030 0.620  3.045 0.562
3.060 0.504  3.075 0.448  3.090 0.394  3.105 0.341
3.120 0.291  3.135 0.244  3.150 0.200  3.165 0.159
3.180 0.123  3.195 0.091  3.210 0.063  3.225 0.040
3.240 0.022  3.255 0.010  3.270 0.002  3.285 0.000
3.300 0.003  /
\putrule from 0.0 0.0 to 3.3 0.0
\putrule from 0.0 -0.1 to 0.0 1.4
\putrule from 0.3 -0.02 to 0.3 0.02
\putrule from 0.6 -0.02 to 0.6 0.02
\putrule from 0.9 -0.02 to 0.9 0.02
\putrule from 1.2 -0.02 to 1.2 0.02
\putrule from 1.5 -0.035 to 1.5 0.035
\putrule from 1.8 -0.02 to 1.8 0.02
\putrule from 2.1 -0.02 to 2.1 0.02
\putrule from 2.4 -0.02 to 2.4 0.02
\putrule from 2.7 -0.02 to 2.7 0.02
\putrule from 3.0 -0.035 to 3.0 0.035
\putrule from -0.035 1.0 to 0.035 1.0
\putrule from -0.035 0.5 to 0.035 0.5
\put {$A(y)$} [c] at 0.15 1.4 
\put {$y$} [c] at 3.3 0.15 
\put {$0$} [c] at -0.15 0 
\put {$0.5$} [c] at 1.5 -0.15 
\put {$1.0$} [c] at 3.0 -0.15
\put {$0.1$} [c] at -0.15 1.00 
\endpicture$$
{\ninepoint  \baselineskip=10pt
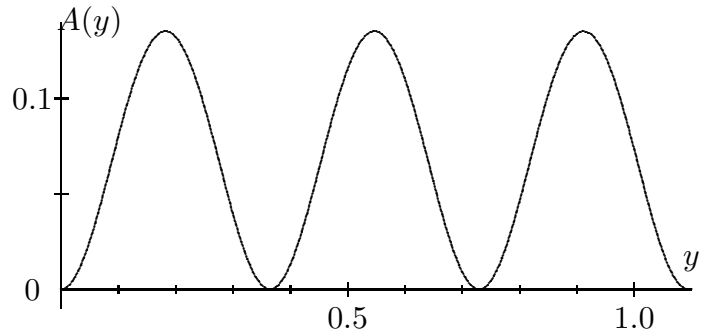
{\bf Figure 2.\/}  A periodic warp function $A(y)$ for $\Lambda=-1$,
$\lambda=0$ and $\mu=-0.01$.  The initial condition is $A^{\prime\prime}(0) =
21.460889$.  The value of $k_0(\phi')^2$ oscillates  about $-4$, also with a
small period and amplitude.}
\medskip
%

We can find numerical solutions arbitrarily close to the surface $\mu=0$, for
$\lambda<{9\over 4}$ and $\Lambda=-1$, and to the surface described by
$\spsolncmore$, provided $\mu > -{81\over 128}$.  Apparently, another boundary
exists for solutions with $\mu < -{81\over 128}$ which extends approximately
along the line $\mu\sim {9\over 8}\lambda$.  We have indicated this boundary
in figure 1 by extending the shaded region beyond the curve $\spsolncmore$. 
As $\lambda$ and $\mu$ approach this boundary from above, the numerical
solutions grow more cusped as in the more extreme example depicted in figure 3
where the warp function has broad maxima and sharp, deep minima.  The
amplitude is also significantly larger so that $e^{A(y)}$ changes by several
orders of magnitude.  Although the figure appears to have cusps at the minima,
the function $A(y)$ is actually smooth at these points and has no
discontinuities in the slope.  
$$\beginpicture
\setcoordinatesystem units <1.0truein,1.0truein>
\setplotarea x from 0.0 to 3.5, y from -0.1 to 1.4
\setlinear
\plot  0.000 0.000  .0050 .0411  .0100 .1326  .0150 .2293
.0200 .3137  .0250 .3840  .0300 .4428  .0350 .4927
.0400 .5355  .0450 .5728  .0500 .6058  
.0625 .6738  .0750 .7277  .0875 .7717
.1000 .8088  .1125 .8407  .1250 .8685  .1375 .8934
.1500 .9154  .1625 .9354  .1750 .9534  .1875 .9699
.2000 .9852  .2125 .9992  .2250 1.012  .2375 1.024
.2500 1.036  .2625 1.046  .2750 1.056  .2875 1.065
.3000 1.074  .3125 1.082  .3250 1.090  .3375 1.097
.3500 1.104  .3625 1.110  .3750 1.116  .3875 1.122
.4000 1.128  .4125 1.133  .4250 1.138  .4375 1.142
.4500 1.147  .4625 1.151  .4750 1.155  .4875 1.158
.5000 1.162  .5125 1.165  .5250 1.168  .5375 1.171
.5500 1.174  .5625 1.176  .5750 1.178  .5875 1.180
.6000 1.182  .6125 1.184  .6250 1.186  .6375 1.187
.6500 1.188  .6625 1.190  .6750 1.191  .6875 1.191
.7000 1.192  .7125 1.193  .7875 1.193  .8000 1.192
.8125 1.192  .8250 1.191  .8375 1.190  .8500 1.189
.8625 1.187  .8750 1.186  .8875 1.184  .9000 1.183
.9125 1.181  .9250 1.179  .9375 1.176  .9500 1.174
.9625 1.171  .9750 1.169  .9875 1.166  1.000 1.162
1.012 1.159  1.025 1.155  1.038 1.152  1.050 1.147
1.062 1.143  1.075 1.139  1.088 1.134  1.100 1.129
1.112 1.123  1.125 1.117  1.138 1.111  1.150 1.105
1.162 1.098  1.175 1.091  1.188 1.083  1.200 1.075
1.212 1.067  1.225 1.058  1.238 1.048  1.250 1.037
1.262 1.026  1.275 1.014  1.288 1.002  1.300 .9877
1.312 .9729  1.325 .9565  1.338 .9387  1.350 .9192
1.362 .8975  1.375 .8734  1.388 .8462  1.400 .8150
1.412 .7790  1.425 .7363  1.438 .6847  1.450 .6198
1.462 .5342  1.475 .4128  1.488 .2265  1.490 .1792
1.492 .1298  1.495 .0814  1.498 .0391  1.500 .0099
1.502 .0002  1.505 .0123  1.508 .0432  1.510 .0864
1.512 .1353  1.525 .3522  1.538 .4938  1.550 .5905
1.562 .6622  1.575 .7182  1.588 .7638  1.600 .8020
1.612 .8349  1.625 .8635  1.638 .8887  1.650 .9112
1.662 .9315  1.675 .9500  1.688 .9669  1.700 .9822
1.712 .9965  1.725 1.010  1.738 1.022  1.750 1.033
1.762 1.044  1.775 1.054  1.788 1.063  1.800 1.072
1.812 1.080  1.825 1.088  1.838 1.096  1.850 1.102
1.862 1.109  1.875 1.115  1.888 1.121  1.900 1.127
1.912 1.132  1.925 1.137  1.938 1.141  1.950 1.146
1.962 1.150  1.975 1.154  1.988 1.158  2.000 1.161
2.012 1.164  2.025 1.167  2.038 1.170  2.050 1.173
2.062 1.175  2.075 1.178  2.088 1.180  2.100 1.182
2.112 1.184  2.125 1.185  2.138 1.187  2.150 1.188
2.162 1.189  2.175 1.190  2.188 1.191  2.200 1.192
2.212 1.192  2.225 1.193  2.288 1.193  2.300 1.192
2.312 1.192  2.325 1.191  2.338 1.190  2.350 1.189
2.362 1.188  2.375 1.186  2.388 1.185  2.400 1.183
2.412 1.181  2.425 1.179  2.438 1.177  2.450 1.174
2.462 1.172  2.475 1.169  2.488 1.166  2.500 1.163
2.512 1.160  2.525 1.156  2.538 1.152  2.550 1.148
2.562 1.144  2.575 1.139  2.588 1.135  2.600 1.130
2.612 1.124  2.625 1.119  2.638 1.113  2.650 1.106
2.662 1.100  2.675 1.092  2.688 1.085  2.700 1.077
2.712 1.068  2.725 1.059  2.738 1.050  2.750 1.040
2.762 1.029  2.775 1.017  2.788 1.004  2.800 .9905
2.812 .9759  2.825 .9599  2.838 .9424  2.850 .9232
2.862 .9019  2.875 .8782  2.888 .8517  2.900 .8213
2.912 .7863  2.925 .7452  2.938 .6955  2.950 .6335
2.962 .5527  2.975 .4397  2.980 .3803  2.985 .3092
2.990 .2240  2.995 .1268  3.000 .0366  3.005 .0002
3.010 .0449  3.015 .1376  3.020 .2338  3.025 .3173
3.038 .4715  3.050 .5748
3.062 .6502  3.075 .7087  3.088 .7560  3.100 .7955
3.112 .8292  3.125 .8585  3.138 .8844  3.150 .9074
3.162 .9280  3.175 .9467  3.188 .9639  3.200 .9795
3.212 .9940  3.225 1.007  3.238 1.020  3.250 1.031
3.262 1.042  3.275 1.052  3.288 1.062  3.300 1.071
3.312 1.079  3.325 1.087  3.338 1.094  3.350 1.101
3.362 1.108  3.375 1.114  3.388 1.120  3.400 1.126
3.412 1.131  3.425 1.136  3.438 1.141  3.450 1.145
3.462 1.149  3.475 1.153  3.488 1.157  3.500 1.161  /
\putrule from 0.0 0.0 to 3.5 0.0
\putrule from 0.0 -0.1 to 0.0 1.4
\putrule from 0.5 -0.035 to 0.5 0.035
\putrule from 1.0 -0.035 to 1.0 0.035
\putrule from 1.5 -0.035 to 1.5 0.035
\putrule from 2.0 -0.035 to 2.0 0.035
\putrule from 2.5 -0.035 to 2.5 0.035
\putrule from 3.0 -0.035 to 3.0 0.035
\putrule from -0.035 1.0 to 0.035 1.0
\putrule from -0.035 0.667 to 0.035 0.667
\putrule from -0.035 0.333 to 0.035 0.333
\put {$A(y)$} [c] at 0.15 1.4 
\put {$y$} [c] at 3.5 0.15 
\put {$0$} [c] at -0.15 0 
\put {$1$} [c] at 0.5 -0.15 
\put {$2$} [c] at 1.0 -0.15
\put {$3$} [c] at 1.5 -0.15 
\put {$4$} [c] at 2.0 -0.15
\put {$5$} [c] at 2.5 -0.15 
\put {$6$} [c] at 3.0 -0.15
\put {$6$} [c] at -0.15 1.00 
\put {$4$} [c] at -0.15 0.667
\put {$2$} [c] at -0.15 0.333
\endpicture$$
{\ninepoint  \baselineskip=10pt
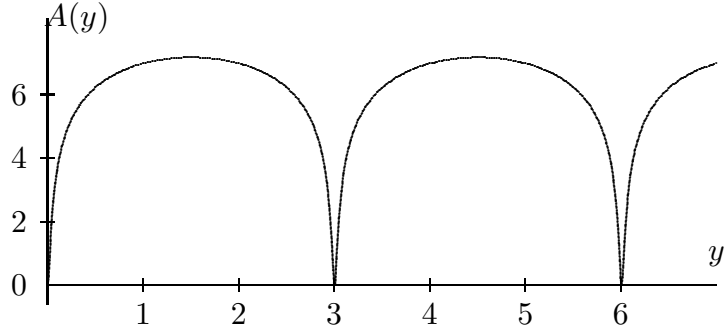
{\bf Figure 3.\/}  A periodic warp function $A(y)$ for $\Lambda = \lambda =
\mu = -1$.  The initial condition is $A^{\prime\prime}(0) = 5367.89$.  Despite
their cusped appearance, the minima are smooth.}
\medskip
%

Most solutions within the unshaded region of figure 1 tend to lie between the
extremes depicted in figures 2 and 3 as in the example provided by the
following sketch.  More examples are presented in Appendix B.  
$$\beginpicture
\setcoordinatesystem units <1.0truein,1.0truein>
\setplotarea x from 0.0 to 3.0, y from -0.1 to 1.2
\setlinear
\plot  0.000 0.000  .0150 .0141  .0300 .0539  .0450 .1133
.0600 .1841  .0750 .2591  .0900 .3330  .1050 .4027
.1200 .4670  .1350 .5253  .1500 .5780  .1650 .6253
.1800 .6680  .1950 .7063  .2100 .7410  .2250 .7723
.2400 .8007  .2550 .8267  .2700 .8500  .2850 .8713
.3000 .8910  .3150 .9087  .3300 .9247  .3450 .9397
.3600 .9530  .3750 .9653  .3900 .9763  .4050 .9863
.4200 .9953  .4350 1.004  .4500 1.011  .4650 1.017
.4800 1.023  .4950 1.028  .5100 1.032  .5250 1.035
.5400 1.038  .5550 1.039  .5700 1.041  .5850 1.041
.6000 1.041  .6150 1.040  .6300 1.039  .6450 1.037
.6600 1.034  .6750 1.030  .6900 1.026  .7050 1.021
.7200 1.015  .7350 1.008  .7500 1.001  .7650 .9923
.7800 .9830  .7950 .9723  .8100 .9610  .8250 .9483
.8400 .9343  .8550 .9190  .8700 .9023  .8850 .8840
.9000 .8640  .9150 .8417  .9300 .8173  .9450 .7907
.9600 .7613  .9750 .7287  .9900 .6927  1.005 .6530
1.020 .6087  1.035 .5593  1.050 .5047  1.065 .4443
1.080 .3780  1.095 .3065  1.110 .2317  1.125 .1575
1.140 .0899  1.155 .0368  1.170 .0057  1.179 .0000
1.185 .0019  1.200 .0258  1.215 .0737  1.230 .1381
1.245 .2112  1.260 .2863  1.275 .3590  1.290 .4267
1.305 .4887  1.320 .5450  1.335 .5957  1.350 .6413
1.365 .6823  1.380 .7193  1.395 .7527  1.410 .7830
1.425 .8103  1.440 .8353  1.455 .8580  1.470 .8787
1.485 .8977  1.500 .9147  1.515 .9303  1.530 .9447
1.545 .9577  1.560 .9693  1.575 .9800  1.590 .9897
1.605 .9983  1.620 1.006  1.635 1.013  1.650 1.019
1.665 1.025  1.680 1.029  1.695 1.033  1.710 1.036
1.725 1.038  1.740 1.040  1.755 1.041  1.770 1.041
1.785 1.041  1.800 1.040  1.815 1.038  1.830 1.036
1.845 1.033  1.860 1.029  1.875 1.024  1.890 1.019
1.905 1.013  1.920 1.006  1.935 .9977  1.950 .9890
1.965 .9793  1.980 .9683  1.995 .9563  2.010 .9433
2.025 .9290  2.040 .9133  2.055 .8960  2.070 .8770
2.085 .8560  2.100 .8330  2.115 .8080  2.130 .7803
2.145 .7497  2.160 .7160  2.175 .6787  2.190 .6373
2.205 .5913  2.220 .5400  2.235 .4833  2.250 .4207
2.265 .3523  2.280 .2795  2.295 .2044  2.310 .1317
2.325 .0685  2.340 .0226  2.355 .0011  2.359 .0000
2.370 .0075  2.385 .0408  2.400 .0956  2.415 .1641
2.430 .2386  2.445 .3132  2.460 .3843  2.475 .4500
2.490 .5100  2.505 .5640  2.520 .6127  2.535 .6567
2.550 .6963  2.565 .7320  2.580 .7640  2.595 .7933
2.610 .8197  2.625 .8440  2.640 .8657  2.655 .8857
2.670 .9040  2.685 .9207  2.700 .9357  2.715 .9493
2.730 .9620  2.745 .9733  2.760 .9837  2.775 .9930
2.790 1.001  2.805 1.009  2.820 1.016  2.835 1.021
2.850 1.026  2.865 1.031  2.880 1.034  2.895 1.037
2.910 1.039  2.925 1.040  2.940 1.041  2.955 1.041
2.970 1.041  2.985 1.039  3.000 1.037  /
\putrule from 0.0 0.0 to 3.0 0.0
\putrule from 0.0 -0.1 to 0.0 1.1
\putrule from 0.5 -0.035 to 0.5 0.035
\putrule from 1.0 -0.035 to 1.0 0.035
\putrule from 1.5 -0.035 to 1.5 0.035
\putrule from 2.0 -0.035 to 2.0 0.035
\putrule from 2.5 -0.035 to 2.5 0.035
\putrule from 3.0 -0.035 to 3.0 0.035
\putrule from -0.035 1.0 to 0.035 1.0
\putrule from -0.035 0.667 to 0.035 0.667
\putrule from -0.035 0.333 to 0.035 0.333
\put {$A(y)$} [c] at 0.15 1.2 
\put {$y$} [c] at 3.0 0.15 
\put {$0$} [c] at -0.15 0 
\put {$1$} [c] at 0.333 -0.15 
\put {$2$} [c] at 0.667 -0.15 
\put {$3$} [c] at 1.0 -0.15
\put {$4$} [c] at 1.333 -0.15 
\put {$5$} [c] at 1.667 -0.15 
\put {$6$} [c] at 2.0 -0.15
\put {$7$} [c] at 2.333 -0.15 
\put {$8$} [c] at 2.667 -0.15 
\put {$9$} [c] at 3.0 -0.15
\put {$3$} [c] at -0.15 1.00 
\put {$2$} [c] at -0.15 0.667
\put {$1$} [c] at -0.15 0.333
\endpicture$$
{\ninepoint  \baselineskip=10pt
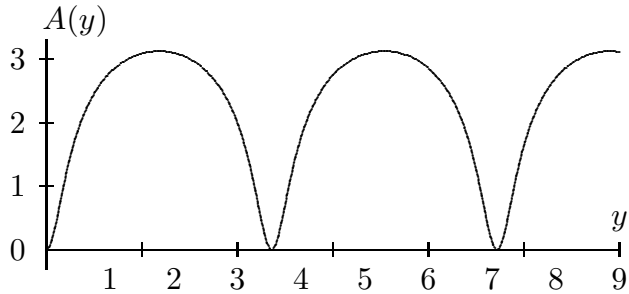
{\bf Figure 4.\/}  An example of a periodic warp function that lies between
the extremes represented by figures 2 and 3.  For this sketch we used $\Lambda
= -1$, $\lambda=-3$ and $\mu = -2$.  The initial condition is
$A^{\prime\prime}(0) = 42.2832125$.  }
\medskip
%

Finally, we must address whether the kinetic energy term in the scalar action
can have the correct sign, $k_0=1$, when $A(y)$ is periodic.  If we include
the additional term\foot{Other terms of the same order are possible, such as 
$R_{ab}\nabla^a\phi\nabla^b\phi$, but the squared kinetic term is the simplest
to analyze.} 
$$-{1\over 4} k_1 (\nabla_a\phi\nabla^a\phi)^2 \neqno\sqkin$$
then the field equations $\eom$ require 
$$\eqalign{
\mu\left[ \half A^{\prime\prime\prime\prime} + 2 A' A^{\prime\prime\prime} +
{\textstyle{3\over 2}} (A^{\prime\prime})^2 + 2 A^{\prime\prime} (A')^2
\right] & \cr
+ {\textstyle{3\over 4}} \lambda \left[ 2 A^{\prime\prime} (A')^2 + (A')^4
\right] - {\textstyle{3\over 2}} (A')^2 - {\textstyle{3\over 2}}
A^{\prime\prime} &= - \Lambda + {\textstyle{1\over 4}} k_0 (\phi')^2 +
{\textstyle{1\over 8}} k_1 (\phi')^4 \cr
\mu\left[ A' A^{\prime\prime\prime} - \half (A^{\prime\prime})^2 + 2
A^{\prime\prime} (A')^2 \right]  & \cr
+ {\textstyle{3\over 4}} \lambda (A')^4 - {\textstyle{3\over 2}} (A')^2 
&= - \Lambda - {\textstyle{1\over 4}} k_0 (\phi')^2 - {\textstyle{3\over 8}}
k_1 (\phi')^4 . \cr}
\neqno\bothkineqns$$
Since these expressions are only quadratic in $(\phi'(y))^2$, we can solve for 
$$\eqalign{(\phi'(y))^2 
&= {k_0\over 3k_1} \biggl[ -1 \pm 
\biggl\{ 1 -  24 {k_1\over k_0^2} \biggl[ \Lambda + 
\mu\left( A'A^{\prime\prime\prime} - {1\over 2} (A^{\prime\prime})^2 + 2
A^{\prime\prime}(A')^2 \right) \cr
&\qquad\qquad\qquad
+ {3\over 4} \lambda (A')^4 - {3\over 2} (A')^2 \biggr] \biggr\}^{1/2} \biggr]
\cr}
\neqno\phisqeqn$$
and substitute the result into $\bothkineqns$ to obtain a differential
equation for $A(y)$.  This equation still admits periodic solutions such as
the example sketched in figure 5 for $\Lambda=-1$, $\lambda=0$, $\mu=-0.1$,
$k_0=1$, and $k_1=-0.25$ and choosing the minus root\foot{The choice of the
plus root gives a periodic solution that again requires $k_0<0$.} in
$\phisqeqn$.  The value of $k_0(\phi')^2$ is positive so we obtain the
standard normalization for the scalar kinetic energy.  
$$\beginpicture
\setcoordinatesystem units <1.0truein,1.0truein>
\setplotarea x from 0.0 to 3.0, y from -0.1 to 1.55
\setlinear
\setlinear
\putrule from 0.0 0.0 to 3.0 0.0
\putrule from 0.0 -0.1 to 0.0 1.55
\putrule from 0.5 -0.035 to 0.5 0.035
\putrule from 1.0 -0.035 to 1.0 0.035
\putrule from 1.5 -0.035 to 1.5 0.035
\putrule from 2.0 -0.035 to 2.0 0.035
\putrule from 2.5 -0.035 to 2.5 0.035
\putrule from 3.0 -0.035 to 3.0 0.035
\putrule from -0.035 1.5 to 0.035 1.5
\putrule from -0.035 1.0 to 0.035 1.0
\putrule from -0.035 0.5 to 0.035 0.5
\put {$A(y)$} [l] at 3.05 0.4044
\put {$\phi'(y)$} [l] at 3.05 1.017
\put {$y$} [c] at 3.0 0.15 
\put {$0$} [c] at -0.15 0 
\put {$1$} [c] at 1.0 -0.15
\put {$2$} [c] at 2.0 -0.15
\put {$3$} [c] at 3.0 -0.15
\put {$3$} [c] at -0.15 1.5 
\put {$2$} [c] at -0.15 1.0
\put {$1$} [c] at -0.15 0.5
%
%
\plot 0.0 0.000  .03 .0053  .06 .0208  .09 .0452
.12 .0769  .15 .1137  .18 .1540  .21 .1960
.24 .2384  .27 .2802  .30 .3206  .33 .3588
.36 .3946  .39 .4274  .42 .4572  .45 .4838
.48 .5070  .51 .5265  .54 .5425  .57 .5550
.60 .5640  .63 .5695  .66 .5710  .69 .5690
.72 .5630  .75 .5535  .78 .5405  .81 .5240
.84 .5040  .87 .4801  .90 .4531  .93 .4228
.96 .3896  .99 .3534  1.02 .3148  1.05 .2742
1.08 .2323  1.11 .1898  1.14 .1480  1.17 .1081
1.20 .0719  1.23 .0412  1.26 .0180  1.29 .0039
1.32 .0000  1.35 .0070  1.38 .0238  1.41 .0494
1.44 .0819  1.47 .1194  1.50 .1600  1.53 .2022
1.56 .2446  1.59 .2862  1.62 .3262  1.65 .3642
1.68 .3995  1.71 .4320  1.74 .4613  1.77 .4874
1.80 .5100  1.83 .5290  1.86 .5445  1.89 .5565
1.92 .5650  1.95 .5695  1.98 .5710  2.01 .5680
2.04 .5620  2.07 .5520  2.10 .5385  2.13 .5210
2.16 .5005  2.19 .4764  2.22 .4489  2.25 .4182
2.28 .3844  2.31 .3480  2.34 .3090  2.37 .2682
2.40 .2262  2.43 .1837  2.46 .1420  2.49 .1026
2.52 .0670  2.55 .0373  2.58 .0153  2.61 .0027
2.63 .0000  2.67 .0088  2.70 .0270  2.73 .0538
2.76 .0871  2.79 .1252  2.82 .1661  2.85 .2084
2.88 .2507  2.91 .2922  2.94 .3319  2.97 .3694
3.00 .4044  /
%
%
\setdashes
\setdashpattern <2pt, 2pt>
\plot 0.0 1.518  .03 1.509  .06 1.483  .09 1.444
.12 1.394  .15 1.340  .18 1.285  .21 1.231
.24 1.180  .27 1.134  .30 1.093  .33 1.057
.36 1.026  .39 .9985  .42 .9750  .45 .9560
.48 .9400  .51 .9270  .54 .9170  .57 .9095
.60 .9040  .63 .9010  .66 .9000  .69 .9015
.72 .9050  .75 .9105  .78 .9185  .81 .9290
.84 .9420  .87 .9585  .90 .9785  .93 1.002
.96 1.030  .99 1.062  1.02 1.099  1.05 1.141
1.08 1.188  1.11 1.239  1.14 1.293  1.17 1.349
1.20 1.402  1.23 1.450  1.26 1.488  1.29 1.511
1.32 1.518  1.35 1.506  1.38 1.478  1.41 1.438
1.44 1.387  1.47 1.332  1.50 1.278  1.53 1.224
1.56 1.174  1.59 1.128  1.62 1.088  1.65 1.052
1.68 1.021  1.71 .9945  1.74 .9720  1.77 .9535
1.80 .9380  1.83 .9255  1.86 .9155  1.89 .9085
1.92 .9035  1.95 .9010  1.98 .9000  2.01 .9015
2.04 .9055  2.07 .9115  2.10 .9200  2.13 .9305
2.16 .9445  2.19 .9610  2.22 .9815  2.25 1.006
2.28 1.034  2.31 1.067  2.34 1.104  2.37 1.148
2.40 1.195  2.43 1.246  2.46 1.301  2.49 1.356
2.52 1.410  2.55 1.456  2.58 1.492  2.61 1.514
2.64 1.518  2.67 1.504  2.70 1.473  2.73 1.430
2.76 1.380  2.79 1.324  2.82 1.269  2.85 1.216
2.88 1.167  2.91 1.122  2.94 1.082  2.97 1.048
3.00 1.017  /
\endpicture$$
{\ninepoint  \baselineskip=10pt
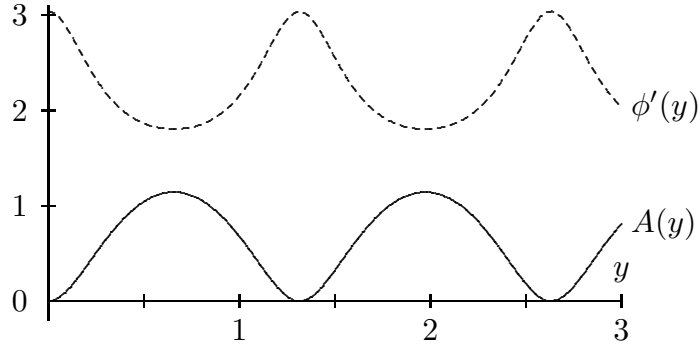
{\bf Figure 5.\/}  A periodic warp function $A(y)$ (solid line) and $\phi'(y)$
(dashed line) for $\Lambda = -1$, $\lambda = 0$, $\mu = -0.1$, $k_0=1$, and
$k_1=-0.25$.  The initial condition is $A^{\prime\prime}(0) = 23.77364592$.  }
\medskip
%

\newsec{Discussion.}

\subsec{Higher Order Terms in the Effective Action.}

The gravitational action that we have considered should be regarded as only
the first few terms of a possibly infinite effective action arranged in powers
of derivatives.  Therefore, we might worry whether higher order terms will
spoil the periodic behavior seen in the last section.  Here we briefly
motivate why the existence of smooth, periodic, non-singular solutions might
be a generic feature of this scenario.  

The small $\Lambda$ region of the generalized parameter space, formed by the
coefficients of the terms in the effective action, still should admit warp
functions that approximate sinusoids as in $\cosapprox$,
$$A(y) = A_0 + \epsilon\cos(\omega(y-y_0)) + {\cal O}(\epsilon^2) .
\neqno\gencosapprox$$
For this approximate solution, the amplitude of the oscillations is assumed to
be small, ${\cal O}(\epsilon)$, so terms in the equations of motion with fewer
powers of $A(y)$ tend to dominate the shape of the warp function.  For an
action with a general set of $R^k$ terms included, the leading behavior in the
$\epsilon\ll 1$ limit is 
$$- {3\over 2} {d^2A\over dy^2} + {1\over 2} \mu {d^4A\over dy^4} +
\sum_{k=3}^\infty \sum_i \mu_{k,i} {d^{2k}A\over dy^{2k}} = {\cal
O}(\epsilon^2) , \neqno\genlinear$$
where the $\mu_{k,i}$ are some linear combinations of the coefficients of the
$R^k$ terms.  When the warp function is of the form $\gencosapprox$, these
linear terms have solutions as long as the coefficients $\mu_{k,i}$ are such
that the equation  
$${3\over 2} + {1\over 2} \mu \omega^2 + \sum_{k=3}^\infty \sum_i
(-1)^k\mu_{k,i} \omega^{2(k-1)} = 0  \neqno\genlinear$$
has a real root.  As before, the cosmological constant can be related to the
amplitude $\epsilon$, by solving for the ${\cal O}(\epsilon^2)$ terms of the
field equations of the full Lagrangian.  Thus, when the cosmological constant
is small, $\Lambda \sim {\cal O}(\epsilon^2)$, then the full parameter space
contains a region in which the warp function is, up to ${\cal O}(\epsilon^2)$
corrections, given by a sinusoid $\gencosapprox$.  As with the parameter
$\lambda$ in the $R^2$ action, the other parameters which appear in the full
equations of motion, which we denote by $\lambda_{k,i}$, only appear in terms
that are at least cubic in powers of the warp function and so only need to
satisfy the weaker requirement that they do not ruin the $\epsilon$ expansion,
$\lambda_{k,i}\Lambda\not\gg 1$.  

The problem of establishing the existence of more general periodic solutions
becomes only more difficult at higher orders.  However, well-behaved periodic
solutions do seem generically to exist for subsets of the terms of the general
field equations other than the linear terms.  As an example, one such subset
of the terms in $\aeqn$,  
$$\half\mu A^{\prime\prime\prime\prime} + \left[ 4\mu + {\textstyle{3\over 2}}
\lambda \right] A^{\prime\prime} (A')^2  = 0  , \neqno\nonlinearega$$
also has periodic solutions when ${\lambda\over\mu} > - {8\over 3}$.  One
subset of the terms in the $R^3$ action is 
$$A^{\prime\prime\prime\prime\prime\prime} + c_1\,
A^{\prime\prime\prime\prime} (A')^2 - c_2 A^{\prime\prime} (A')^4 = 0 ,
\neqno\nonlinearega$$
which has periodic solutions for some subregion of the space $c_1, c_2 > 0$. 
This behavior also suggests that it is plausible that periodic solutions
should exist in some region of the enlarged parameter space when higher order
terms are included in the gravitational action.

\subsec{The Scenario in $d$ Dimensions.}

Although for clarity we have concentrated on a scenario in which the universe
contains $3+1$ infinite dimensions and one extra compact dimension, the
picture can be generalized to higher dimensions with many of the features
found above intact.  The most trivial modification of the metric $\rsmetric$
would be of the form 
$$ds^2 = e^{A(y)}\left[ \eta_{\mu\nu}\, dx^\mu dx^\nu + \sum_{i=1}^{d-5}
dz_i^2 \right] + dy^2  \neqno\multiperiod$$
for a theory in $d$ total space-time dimensions.  Although the warp function
$A(y)$ must still be a periodic function of $y$, the $z_i$ directions can be
trivially compactified with arbitrary compactification radii.  The generalized
Einstein equation $\eom$ for this scenario produces the following two
independent differential equations,
$$\eqalign{
\tilde\mu\, \left[ \half A^{\prime\prime\prime\prime} + {\textstyle{d-1\over
2}} A' A^{\prime\prime\prime} + {\textstyle{3(d-1)\over 8}}
(A^{\prime\prime})^2 + {\textstyle{(d-1)^2\over 8}} A^{\prime\prime} (A')^2
\right] & \cr
+ \tilde\lambda\, \left[ A^{\prime\prime} (A')^2 + {\textstyle{d-1\over 8}}
(A')^4 \right] - {\textstyle{(d-1)(d-2)\over 8}} (A')^2 - {\textstyle{d-2\over
2}} A^{\prime\prime} &= - \Lambda + {\textstyle{1\over 4}} k_0 (\phi')^2 \cr
{\textstyle{d-1\over 4}} \tilde\mu\, \left[ A' A^{\prime\prime\prime} -
{\textstyle{1\over 2}} (A^{\prime\prime})^2 + {\textstyle{d-1\over 2}}
A^{\prime\prime} (A')^2 \right]  & \cr
+ {\textstyle{d-1\over 8}} \tilde\lambda\, (A')^4 -
{\textstyle{(d-1)(d-2)\over 8}} (A')^2 
&= - \Lambda - {\textstyle{1\over 4}} k_0 (\phi')^2 . \cr}
\neqno\geneom$$
As before, only two of the three possible linear combinations of $R^2$ terms
contribute,\foot{Note that in $\mldefs$ we have defined $\lambda = {2\over
3}\tilde\lambda|_{d=5}$ to agree with the conventional normalization for the
Gauss-Bonnet term.}
$$\tilde\mu=4(d-1)\, a + d\, b + 4\, c 
\qquad 
\tilde\lambda = {d-4\over 4} \left[ d(d-1)\, a + (d-1)\, b + 2\, c \right]
.\neqno\gendefs$$
As in the $d=5$ case, for a conformally flat metric such as $\multiperiod$ the
squared Weyl term $C_{abcd}C^{abcd}$ does not contribute to the equations of
motion $\geneom$.  $\tilde\lambda$ vanishes for $d=4$ since the Gauss-Bonnet
term then corresponds to the Euler form and is a topological density. 
Incidentally, for the solution to be periodic in this unphysical case which
only has $2+1$ dimensional Poincar\' e invariance, $\mu$ must be moderately
fine-tuned to lie within the interval $-{1\over 24}<\mu<0$.

An interesting feature of this more general field equation is that the
analytic solutions of $\expspsoln$ are still solutions when $d>5$.  The
generalization of $\spsolnc$ is then
$$\eqalign{
\Lambda &\equiv {(d-1)(d-2)^2\over 36} \bar\Lambda = {b\over 8}
{(d-1)(d-2)\over 2a+d-1} \cr 
\tilde\lambda &\equiv {3\over 2}\bar\lambda = - {a\over b} {(d-2)(3a+d-1)\over
2a+d-1} \cr
\tilde\mu &\equiv {16\bar\mu\over(d-1)^2} = {d-2\over b(2a+d-1)} . \cr}
\neqno\genspsolnc$$
Here we have introduced a rescaled set of parameters, $\{\bar\Lambda,
\bar\lambda, \bar\mu\}$, to emphasize that the shape of the curve shown in
figure 1 $\spsolncmore$ is in fact universal---it is the same for arbitrary
$\Lambda<0$ and in arbitrary dimension $d>4$ when $\mu$ and $\lambda$ are
appropriately rescaled:
$$\bar\lambda\bar\Lambda = - {8\over 3}\, \bar\mu\bar\Lambda \pm 4
\sqrt{2\bar\mu\bar\Lambda} - {9\over 4} . \neqno\universalcurve$$
Note that while the location of this set of solutions is independent of $d$
and $\Lambda$ when plotted in the $\{ \bar\lambda\bar\Lambda,
\bar\mu\bar\Lambda \}$-plane, individual points along this curve do not
correspond to the same solutions since both the period and the exponent in
$\expspsoln$ depend on $d$.

\newsec{Conclusions.}

A theory with an extra compact dimension and an action with a set of $R^2$
terms and a compact scalar field contains sufficient freedom to permit a
metric that maintains Poincar\' e invariance in $3+1$ of the dimensions
without any fine-tuning of the terms in the action.  Although the resulting
field equations are fourth order, the existence of these metrics can be shown
numerically.  The examples that we have found are smooth, periodic and contain
no singularities or zeros throughout space-time.  As a check, we have
confirmed that when the numerical solutions for the warp function and scalar
field are substituted back into the action, the integral over the extra
dimensions gives zero, so that the effective four dimensional cosmological
constant vanishes.  We have also found several exact solutions which, while
they do not themselves provide satisfactory metrics, bound the region of
parameter space in which exist the desired smooth, non-singular metrics that
are periodic in the extra dimension.  Moreover, these exact solutions continue
to exist for an arbitrary number of extra dimensions and can be expressed in a
universal, dimension-independent form.  

As in the original proposal by Rubakov and Shaposhnikov $\rubakov$ to address
the cosmological constant problem using extra dimensions, the existence of
metrics with periodic warp functions that are flat in the other $3+1$
dimensions leaves unanswered the question of whether they are preferred over
other possibilities.  For example, we have also found a class of metrics that,
while still periodic in one dimension, correspond to $3+1$ de Sitter or
anti-de Sitter spaces in the other components.  There is also the question of
stability, and whether these solutions can be reached from generic initial
conditions.  The answers to these problems are central to the search for a
realistic cosmology, but the problems are dynamical in nature and a fine
turning of fundamental parameters is now not obviously required.

\bigskip\bigskip
\centerline{ {\bf Acknowledgements} }

H.~C.~has benefited from discussions with T.~Chow.  This work was supported in
part by Natural Sciences and Engineering Research Council of Canada.

\appendix{A}{Singularities in the Solutions for a Gauss-Bonnet Action.}

In section 4.2, we mentioned that the warp function $\nomusoln$ for a
Gauss-Bonnet action encounters singularities after only a finite interval in
the extra dimension when $\Lambda<0$.  Moreover, the character of these
singularities depends upon the sign of $\lambda$.  When $\lambda<0$, if we let
$A(y)\to -\infty$ in $\nomusoln$, we discover that this divergence occurs at 
$$\pm (y-y_0) = {\pi\sqrt{3}\over 8} {1\over\sqrt{-\Lambda}} \left[ 
\sqrt{ 1 + {\textstyle{ \sqrt{ 1 - {4\over 3}\lambda\Lambda} }} }
+ \sqrt{ 1 - {\textstyle{ \sqrt{ 1 - {4\over 3}\lambda\Lambda} } }} \right] .
\neqno\singus$$
Note that the sum on the right side is always real, provided $\lambda<0$ and
$\Lambda<0$.  

For $\lambda>0$, another type of singularity occurs since then it becomes
possible to have $A^{\prime\prime}(y)$ diverge when $A'(y)=\pm\lambda^{-1/2}$. 
$\aeqn$ and $\phiequal$ imply that $\sqrt{\lambda}\, A'(y)=\pm\sqrt{1\pm x}$,
so inserting $x=0$ into $\nomusoln$, we see that these singularities also
occur at a finite value of $y$,
$$\pm(y-y_0) = 
+ {\textstyle{1\over 2}} {\sqrt{\textstyle{\lambda\over x_0+1}}}
\arctanh{\textstyle{1\over\sqrt{x_0+1}}} 
- {\textstyle{1\over 2}} {\sqrt{\textstyle{\lambda\over x_0-1}}}
\arctan{\textstyle{1\over\sqrt{x_0-1}}}  .
\neqno\singsddA$$

Since $A^{\prime\prime}(y)={4\over 3}\Lambda$ when $y$ is an extremum, when
$\Lambda>0$ the warp function only contains minima.  Thus we have the
possibility that $A(y)\to +\infty$ which, from $\nomusoln$, occurs as
$y\to\pm\infty$.  However, for $\lambda\Lambda > {3\over 4}$,
$A^{\prime\prime}(y)$ diverges at a finite value for $y$.

\appendix{B}{Several More Examples.}

In the following table, we list the properties of a several more examples of
periodic solutions.  
$$\beginpicture
\setcoordinatesystem units <1.0truein,0.225truein>
\setplotarea x from 0 to 2.625, y from -9.0 to 1.0
\setlinear
\putrule from 0 1.5 to 3.5 1.5
\putrule from 0 0.05 to 3.5 0.05
\putrule from 0 -0.05 to 3.5 -0.05
\putrule from 0 -9 to 3.5 -9
\put {$\lambda$} [c] at 0.25 0.75 
\put {$\mu$} [c] at 0.75 0.75 
\put {$A^{\prime\prime}(0)$} [c] at 1.5 0.75 
\put {period} [c] at 2.25 0.75 
\put {$A_{\rm max}-A_{\rm min}$} [c] at 3.0 0.75 
\put {${1\over 10}$} [c] at 0.25 -0.5 
\put {$-{1\over 10}$} [c] at 0.75 -0.5 
\put {$8.38996422$} [c] at 1.50 -0.5 
\put {$1.2596$} [c] at 2.25 -0.5 
\put {$0.5174$} [c] at 3.0 -0.5 
\put {$1$} [c] at 0.25 -1.5 
\put {$-{1\over 30}$} [c] at 0.75 -1.5 
\put {$12.20618213$} [c] at 1.50 -1.5 
\put {$0.9608$} [c] at 2.25 -1.5 
\put {$0.4321$} [c] at 3.0 -1.5 
\put {$0$} [c] at 0.25 -2.5 
\put {$-{1\over 4}$} [c] at 0.75 -2.5 
\put {$8.96881784$} [c] at 1.50 -2.5 
\put {$2.4476$} [c] at 2.25 -2.5 
\put {$1.4186$} [c] at 3.0 -2.5 
\put {$0$} [c] at 0.25 -3.5 
\put {$-{1\over 10}$} [c] at 0.75 -3.5 
\put {$8.35597222$} [c] at 1.50 -3.5 
\put {$1.2264$} [c] at 2.25 -3.5 
\put {$0.4950$} [c] at 3.0 -3.5 
\put {$0$} [c] at 0.25 -4.5 
\put {$-{1\over 1000}$} [c] at 0.75 -4.5 
\put {$64.61423$} [c] at 1.50 -4.5 
\put {$0.1148$} [c] at 2.25 -4.5 
\put {$0.0422$} [c] at 3.0 -4.5 
\put {$-10$} [c] at 0.25 -5.5 
\put {$-1$} [c] at 0.75 -5.5 
\put {$3.7145907$} [c] at 1.50 -5.5 
\put {$2.1898$} [c] at 2.25 -5.5 
\put {$0.6545$} [c] at 3.0 -5.5 
\put {$-1$} [c] at 0.25 -6.5 
\put {$-1$} [c] at 0.75 -6.5 
\put {$5367.88$} [c] at 1.50 -6.5 
\put {$3.005$} [c] at 2.25 -6.5 
\put {$7.159$} [c] at 3.0 -6.5 
\put {$-10$} [c] at 0.25 -7.5 
\put {$-10$} [c] at 0.75 -7.5 
\put {$34282$} [c] at 1.50 -7.5 
\put {$4.419$} [c] at 2.25 -7.5 
\put {$8.501$} [c] at 3.0 -7.5 
\put {$-256$} [c] at 0.25 -8.5 
\put {$-256$} [c] at 0.75 -8.5 
\put {$10439$} [c] at 1.50 -8.5 
\put {$9.524$} [c] at 2.25 -8.5 
\put {$8.633$} [c] at 3.0 -8.5 
\endpicture$$
{\ninepoint  \baselineskip=10pt
{\bf Table 1.\/}  A brief list of some of values of $\lambda$ and $\mu$ that
give periodic, smooth, non-singular solutions when $\Lambda=-1$.  }
\medskip

\noindent
In the table 2, we hold $\lambda$ and $\mu$ fixed while varying $\Lambda$.  
$$\beginpicture
\setcoordinatesystem units <1.0truein,0.225truein>
\setplotarea x from 0.5 to 2.625, y from -8.0 to 1.0
\setlinear
\putrule from 0.5 1.5 to 3.5 1.5
\putrule from 0.5 0.05 to 3.5 0.05
\putrule from 0.5 -0.05 to 3.5 -0.05
\putrule from 0.5 -8 to 3.5 -8
\put {$\Lambda$} [c] at 0.75 0.75 
\put {$A^{\prime\prime}(0)$} [c] at 1.4375 0.75 
\put {period} [c] at 2.25 0.75 
\put {$A_{\rm max}-A_{\rm min}$} [c] at 3.0 0.75 
\put {$-20$} [c] at 0.75 -0.5 
\put {$494520$} [c] at 1.4375 -0.5 
\put {$0.7210$} [c] at 2.25 -0.5 
\put {$8.0777$} [c] at 3.0 -0.5 
\put {$-10$} [c] at 0.75 -1.5 
\put {$53680$} [c] at 1.4375 -1.5 
\put {$0.9502$} [c] at 2.25 -1.5 
\put {$7.1593$} [c] at 3.0 -1.5 
\put {$-4$} [c] at 0.75 -2.5 
\put {$87404$} [c] at 1.4375 -2.5 
\put {$1.9086$} [c] at 2.25 -2.5 
\put {$7.4489$} [c] at 3.0 -2.5 
\put {$-4$} [c] at 0.75 -3.5 
\put {$47.264364$} [c] at 1.4375 -3.5 
\put {$1.4166$} [c] at 2.25 -3.5 
\put {$1.9249$} [c] at 3.0 -3.5 
\put {$-2$} [c] at 0.75 -4.5 
\put {$14.5276037$} [c] at 1.4375 -4.5 
\put {$1.262$} [c] at 2.25 -4.5 
\put {$0.7918$} [c] at 3.0 -4.5 
\put {$-{1\over 2}$} [c] at 0.75 -5.5 
\put {$5.32842445$} [c] at 1.4375 -5.5 
\put {$1.1708$} [c] at 2.25 -5.5 
\put {$0.3155$} [c] at 3.0 -5.5 
\put {$-{1\over 4}$} [c] at 0.75 -6.5 
\put {$3.5543619$} [c] at 1.4375 -6.5 
\put {$1.1586$} [c] at 2.25 -6.5 
\put {$0.2167$} [c] at 3.0 -6.5 
\put {$-{1\over 10}$} [c] at 0.75 -7.5 
\put {$2.14671387$} [c] at 1.4375 -7.5 
\put {$1.1517$} [c] at 2.25 -7.5 
\put {$0.1348$} [c] at 3.0 -7.5 
\endpicture$$
{\ninepoint  \baselineskip=10pt
{\bf Table 2.\/}  A list of values of the cosmological constant $\Lambda$ that
give periodic, smooth, non-singular solutions given $\lambda = \mu = -{1\over
10}$.  }
\medskip

\noindent
Note that a periodic solution for a specific choice of $\{ \Lambda, \lambda,
\mu \}$ is not necessarily unique; two different values for
$A^{\prime\prime}(0)$ can lead to two different periodic solutions.  An
example of this phenomenon is listed in table 2 and illustrated in figure 5.
$$\beginpicture
\setcoordinatesystem units <1.0truein,1.0truein>
\setplotarea x from 0.0 to 4.0, y from -0.1 to 1.6
\setlinear
\plot 0.0 0.000  .01 .3796  .02 .6460  .03 .7972
.04 .8988  .05 .9738  .06 1.033  .07 1.081
.08 1.122  .09 1.156  .10 1.187  .11 1.213
.12 1.237  .13 1.258  .14 1.278  .15 1.295
.16 1.312  .17 1.326  .18 1.340  .19 1.352
.20 1.364  .21 1.375  .22 1.385  .23 1.394
.24 1.403  .25 1.411  .30 1.443  .35 1.465
.40 1.479  .45 1.487  .50 1.490  .55 1.488
.60 1.482  .65 1.473  .70 1.463  .75 1.451
.80 1.440  .85 1.430  .90 1.424  .95 1.421
1.00 1.423  1.05 1.429  1.10 1.438  1.15 1.450
1.20 1.461  1.25 1.472  1.30 1.481  1.35 1.487
1.40 1.490  1.45 1.488  1.50 1.481  1.55 1.468
1.60 1.447  1.65 1.417  1.70 1.373  1.75 1.309
1.75 1.309  1.76 1.293  1.77 1.275  1.78 1.255
1.79 1.234  1.80 1.209  1.81 1.182  1.82 1.151
1.83 1.116  1.84 1.074  1.85 1.024  1.86 .9634
1.87 .8852  1.88 .7782  1.89 .6156  1.90 .3210
1.91 .0000  1.92 .4300  1.93 .6716  1.94 .8132
1.95 .9098  1.96 .9822  1.97 1.039  1.98 1.086
1.99 1.126  2.00 1.160  2.01 1.190  2.02 1.216
2.03 1.240  2.04 1.261  2.05 1.280  2.06 1.297
2.07 1.313  2.08 1.328  2.09 1.341  2.10 1.353
2.11 1.365  2.12 1.376  2.13 1.385  2.14 1.395
2.15 1.403  2.16 1.411  2.17 1.418  2.18 1.425
2.19 1.432  2.20 1.438  2.21 1.443  2.22 1.448
2.23 1.453  2.24 1.457  2.25 1.462  2.30 1.477
2.35 1.486  2.40 1.489  2.45 1.488  2.50 1.483
2.55 1.475  2.60 1.465  2.65 1.453  2.70 1.442
2.75 1.432  2.80 1.424  2.85 1.421  2.90 1.422
2.95 1.427  3.00 1.436  3.05 1.447  3.10 1.458
3.15 1.469  3.20 1.479  3.25 1.486  3.30 1.489
3.35 1.488  3.40 1.482  3.45 1.470  3.50 1.451
3.51 1.446  3.52 1.441  3.53 1.435  3.54 1.429
3.55 1.422  3.56 1.415  3.57 1.408  3.58 1.399
3.59 1.391  3.60 1.381  3.61 1.371  3.62 1.360
3.63 1.348  3.64 1.335  3.65 1.321  3.66 1.306
3.67 1.290  3.68 1.272  3.69 1.252  3.70 1.230
3.71 1.205  3.72 1.178  3.73 1.146  3.74 1.110
3.75 1.067  3.76 1.016  3.77 .9532  3.78 .8716
3.79 .7588  3.80 .5840  3.81 .2568  3.82 .0000
3.83 .4800  3.84 .7002  3.85 .8326  3.86 .9246
3.87 .9942  3.88 1.050  3.89 1.095  3.90 1.134
3.91 1.167  3.92 1.196  3.93 1.222  3.94 1.245
3.95 1.266  3.96 1.284  3.97 1.301  3.98 1.317
3.99 1.331  4.00 1.345  /
\putrule from 0.0 0.0 to 4.0 0.0
\putrule from 0.0 -0.1 to 0.0 1.5
\putrule from 1.0 -0.035 to 1.0 0.035
\putrule from 2.0 -0.035 to 2.0 0.035
\putrule from 3.0 -0.035 to 3.0 0.035
\putrule from 4.0 -0.035 to 4.0 0.035
\putrule from -0.035 1.2 to 0.035 1.2
\putrule from -0.035 0.8 to 0.035 0.8
\putrule from -0.035 0.4 to 0.035 0.4
\put {$A(y)$} [c] at 0.15 1.6 
\put {$y$} [c] at 4.1 0.15 
\put {$0$} [c] at -0.15 0 
\put {$1$} [c] at 1.0 -0.15 
\put {$2$} [c] at 2.0 -0.15
\put {$3$} [c] at 3.0 -0.15 
\put {$4$} [c] at 4.0 -0.15
\put {$6$} [c] at -0.15 1.2
\put {$4$} [c] at -0.15 0.8
\put {$2$} [c] at -0.15 0.4
\setdashes
\setdashpattern <2pt, 2pt>
\plot  0.0 0.000  .02 .0019  .04 .0074  .06 .0164  .08 .0283  .10 .0428
.12 .0591  .14 .0769  .16 .0956  .18 .1148  .20 .1342  .22 .1533
.24 .1721  .26 .1903  .28 .2078  .30 .2246  .32 .2406  .34 .2558
.36 .2700  .38 .2832  .40 .2958  .42 .3074  .44 .3180  .46 .3280
.48 .3370  .50 .3452  .52 .3526  .54 .3592  .56 .3650  .58 .3700
.60 .3744  .62 .3780  .64 .3808  .66 .3828  .68 .3842  .70 .3850
.72 .3848  .74 .3840  .76 .3826  .78 .3804  .80 .3774  .82 .3736
.84 .3692  .86 .3640  .88 .3580  .90 .3514  .92 .3438  .94 .3354
.96 .3262  .98 .3162  1.00 .3054  1.02 .2936  1.04 .2810  1.06 .2676
1.08 .2532  1.10 .2380  1.12 .2218  1.14 .2050  1.16 .1873  1.18 .1689
1.20 .1501  1.22 .1309  1.24 .1115  1.26 .0924  1.28 .0738  1.30 .0562
1.32 .0402  1.34 .0261  1.36 .0146  1.38 .0062  1.40 .0013  1.42 .0000
1.44 .0026  1.46 .0087  1.48 .0182  1.50 .0306  1.52 .0454  1.54 .0621
1.56 .0801  1.58 .0989  1.60 .1181  1.62 .1374  1.64 .1565  1.66 .1752
1.68 .1934  1.70 .2108  1.72 .2274  1.74 .2432  1.76 .2582  1.78 .2722
1.80 .2854  1.82 .2978  1.84 .3092  1.86 .3198  1.88 .3294  1.90 .3384
1.92 .3464  1.94 .3538  1.96 .3602  1.98 .3660  2.00 .3708  2.02 .3750
2.04 .3784  2.06 .3812  2.08 .3832  2.10 .3844  2.12 .3850  2.14 .3848
2.16 .3838  2.18 .3822  2.20 .3798  2.22 .3768  2.24 .3730  2.26 .3684
2.28 .3630  2.30 .3570  2.32 .3502  2.34 .3424  2.36 .3340  2.38 .3246
2.40 .3144  2.42 .3034  2.44 .2916  2.46 .2788  2.48 .2652  2.50 .2506
2.52 .2352  2.54 .2190  2.56 .2020  2.58 .1842  2.60 .1658  2.62 .1468
2.64 .1276  2.66 .1083  2.68 .0892  2.70 .0707  2.72 .0534  2.74 .0376
2.76 .0240  2.78 .0130  2.80 .0051  2.82 .0008  2.84 .0000  2.86 .0034
2.88 .0101  2.90 .0201  2.92 .0330  2.94 .0481  2.96 .0651  2.98 .0832
3.00 .1021  3.02 .1214  3.04 .1407  3.06 .1598  3.08 .1784  3.10 .1964
3.12 .2136  3.14 .2302  3.16 .2458  3.18 .2606  3.20 .2746  3.22 .2876
3.24 .2998  3.26 .3110  3.28 .3214  3.30 .3310  3.32 .3398  3.34 .3478
3.36 .3548  3.38 .3612  3.40 .3668  3.42 .3716  3.44 .3756  3.46 .3790
3.48 .3816  3.50 .3834  3.52 .3846  3.54 .3850  3.56 .3846  3.58 .3836
3.60 .3818  3.62 .3794  3.64 .3762  3.66 .3722  3.68 .3676  3.70 .3622
3.72 .3558  3.74 .3488  3.76 .3410  3.78 .3324  3.80 .3230  3.82 .3126
3.84 .3016  3.86 .2894  3.88 .2766  3.90 .2628  3.92 .2482  3.94 .2326
3.96 .2162  3.98 .1990  4.00 .1811  /
\endpicture$$
{\ninepoint  \baselineskip=10pt
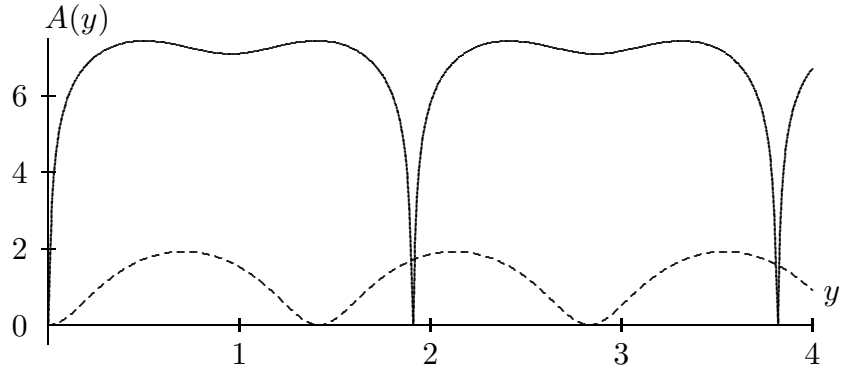
{\bf Figure 6.\/}  A pair of periodic warp function $A(y)$ for $\Lambda = -4$
and $\lambda = \mu = -0.1$.  The initial condition for the solid curve is
$A^{\prime\prime}(0) = 87404$ while that of the dashed curve is
$A^{\prime\prime}(0) = 47.264364$.  Again, both functions are smooth
everywhere.  }
\medskip
%

\listrefs
\bye